\documentclass[twocolumn,showpacs,epsfig,fleqn,nobibnotes]{revtex4}

\usepackage{epsfig}
\begin{document}

\newcommand{\qs}{Q_{\rm sat}}
\newcommand{\qsa}{Q_{\rm sat, A}}
\newcommand{\rr}{\mbox{\boldmath $r$}}
\newcommand{\rrn}{\mbox{$r$}} 
\newcommand{\rp}{\mbox{\boldmath $p$}} 
\newcommand{\rqq}{\mbox{\boldmath $q$}} 
\newcommand{\lsim}{\raisebox{-0.5mm}{$\stackrel{<}{\scriptstyle{\sim}}$}}
\newcommand{\gsim}{\raisebox{-0.5mm}{$\stackrel{>}{\scriptstyle{\sim}}$}}
\def\simge{\mathrel{%
   \rlap{\raise 0.511ex \hbox{$>$}}{\lower 0.511ex \hbox{$\sim$}}}}
\def\simle{\mathrel{
   \rlap{\raise 0.511ex \hbox{$<$}}{\lower 0.511ex \hbox{$\sim$}}}}

\def\pom{{I\!\!P}}

\title{Investigating the role of average color dipole size in  BFKL Pomeron phenomenology}
\author{A.I. Lengyel$^{\,\ddag}$ and M.V.T. Machado$^{\,\star}$}

\affiliation{${}^{\ddag}$Institute of Electron Physics, National
Academy of Sciences of Ukraine, Universitetska 21,
UA-88016 Uzhgorod, Ukraine\\
${}^{\star}$High Energy Physics Phenomenology Group, GFPAE IF-UFRGS, CEP 91501-970, Porto Alegre, Brazil}

%\date{\today}
\begin{abstract}
 Based on the QCD dipole picture of the BFKL Pomeron, we investigate the role played by the saturation scale, $Q_{\mathrm{sat}}$, in obtaining physical values for the affective strong coupling  in phenomenological fits to small-$x$ HERA data. The dependence on this scale appears since the collection of color dipoles characterizing the proton target have average size $1/Q_{\mathrm{sat}}$, which is energy dependent. Physically, this means most of the color dipoles are above but sufficiently close to the border between a saturated and the dilute system. The analysis is first performed in the leading-logs BFKL approach in the saddle-point approximation and it could shed light in further investigations using resummed NLO BFKL kernels. 
\vspace{1pc}
\end{abstract}

\maketitle

\section{Introduction}

The study of the high energy behavior of the observables is an outstanding  issue in perturbative QCD in both phenomenological and theoretical viewpoints.  An important approach encoding all order  $\alpha_s\ln (1/x)$ resummation, which should be dominant at high energies, is the QCD dipole picture \cite{dipole}. As usual, $x$ is the Bjorken variable. It was proven that such approach reproduces the  BFKL evolution \cite{BFKL}. The main process is the onium-onium scattering, that is the reaction between two heavy quark-antiquark states (onia).  In the large $N_c$ limit, the original heavy pair and the further radiated soft gluons due to QCD evolution are represented as a collection of color dipoles. The cross section is then written as a convolution between the density of dipoles in each onium state and the dipole-dipole cross section (scattering via two-gluon exchange). The QCD dipole model can be applied to deep inelastic (DIS) process, assuming that the virtual photon at high virtuality $Q^2$ can be described by an onium. On the other hand, the proton is described by a collection of onia with an average onium radius to be determined from phenomenology. This simple approach describes reasonably well \cite{pheno_early,pheno_new} the experimental results at small-$x$  using a small  number of free parameters. 

Starting from a LO BFKL approach in the saddle-point approximation, an analytical expression is obtained for the proton structure function which correctly describes the energy behavior and the usual scaling violations through the effective anomalous dimension. The rise on $x$ is driven by the hard Pomeron intercept, $\alpha_{\pom}=1 + \omega_{\pom}$, with $\omega_{\pom}=4\,\bar{\alpha}_s \ln 2$ (where $\bar{\alpha}_s=\alpha_s N_c/\pi$). In its original formulation \cite{pheno_early}, the the average number of primary dipoles in the proton was fixed as $n_{\mathrm{eff}}$ and their average transverse diameter $r_0 = 2/Q_0$. The first quantity is absorbed in the overall normalization and the second one is fitted (scales the photon virtuality $Q^2$). The resulting quality of fit is pretty good, but it comes out that the effective strong coupling takes either low values $\alpha_s\simeq 0.07-0.09$. This fact suggested that NLO BFKL correction are necessary, including running coupling. Along these lines, recently a pioneering analysis using resummed NLO BFKL kernels in the saddle-point approximation has been done in Ref. \cite{PRS}. The NLO fits give a qualitatively satisfactory account of the running $\alpha_s$ effect but quantitatively the quality of fit remains sizeably higher than the LO BFKL fit. This feature suggests the investigation of other  proposed theoretical resummation schemes and/or to improve those presented in  Ref. \cite{PRS}. 

In this Letter, we study the possibility to obtain a reasonable quality of fit still using a simple LO BFKL approach with a physically acceptable effective strong coupling. In order to do this, we investigating the role played  by the average transverse size of the color dipoles which collectively constitute the proton. Considering the nonlinear QCD approaches \cite{BK}, it is now well known that the transverse momenta of the partons (gluons) are on average shifted to the saturation scale $Q_{\mathrm{sat}}^2(x)= \Lambda^2\, e^{\lambda\,Y}$ at rapidity $Y=\ln \,(1/x)$ \cite{GBW}. Here, we will suppose that a significant number  of dipoles is in a region above but sufficiently close to the border saturated/dilute system. This allows us to get an estimation of the average size of them. Labeling the size of these dipoles as $r_p$, they are characterized by a density depending  on energy and the transverse size. The average size of these dipoles is then $<\!\!r_p\!\!>\,\propto 1/Q_{\mathrm{sat}}(x)$, which is also the the mean distance between the centers of the neighboring dipole. This fact helps us to write down the probability of finding an onium in the proton as a function of an average onium radius. As a consequence, the previous fixed $Q_0$ is replaced by the scale $Q_{\mathrm{sat}}$ and the hard Pomeron intercept is further enhanced producing a larger effective strong coupling. In what follow, we shortly review the main formulas and perform a phenomenological study using the recent experimental results on the proton structure function at small-$x$.

\section{The QCD dipole picture}

The deep inelastic (DIS) process is a  two-scale problem where the hard scale is given by the photon virtuality and the soft one is associated to the proton typical size. In the color QCD dipole approach, the proton is approximately described by a collection of onia with an unknown average onium radius. Then, the DIS cross section is written as a convolution of the probability of finding an onium in the proton and the photon-onium cross section. This is basically equivalent to the  wave function formulation of the $\gamma^* p$ interaction, where the processes are formulated in terms of the probability distribution of a
$q\bar{q}$ pair in the virtual photon, 
convoluted by the dipole-proton cross section. The latter quantity is described by the convolution of 
the probability distribution of primordial dipoles in the proton times
the dipole-dipole BFKL cross-section \cite{dipole}.

The virtual photon can be described in terms of probability distributions, which are proportional to the well know photon wave functions,
\begin{eqnarray}
\Phi^\gamma_{T,L}(z_{\gamma},r_{\gamma};Q^2) = |\Psi_{T,L} \,(z_{\gamma},r_{\gamma},Q^2)|^2\,,
\label{prob_gamma}
\end{eqnarray}
where $\Phi^\gamma_{T,L}$ are the probability distributions 
of finding a dipole configuration of
transverse size $r_{\gamma}$ at a given $z_{\gamma}$, with the variable 
$z_{\gamma}$ being the photon light-cone momentum
fraction carried by the antiquark of mass $m_f$ and electric charge $e_f$.

The photoabsortion total cross sections reads as \cite{munier_peschanski}
\begin{eqnarray}
\sigma_{tot}^{\gamma^* p} & = &\!\!\!\!\int d^2r_{\gamma} \,dz_{\gamma}
\left[(\Phi^\gamma_{T}+ \Phi^\gamma_{L})\,(z_{\gamma},r_{\gamma},Q^2)\right]\nonumber \\
& \times & \!\!\! \!\! \int d^2r_p \,dz_p\, 
\Phi^p(r_p,z_p)\,\sigma_{dip}(r_{\gamma},r_p;Y)\,,
\label{photoabs_xsec}
\end{eqnarray}
where $\Phi^p(r_p,z_p)$ are the probability distributions of dipoles inside the proton. The dipole-cross section, which encodes the hard Pomeron dynamics, reads as:
\begin{eqnarray}
\sigma_{dip}(r_{\gamma},r_p;Y)=4\pi r_{\gamma}^2 \int\frac{d\gamma}{2i\pi}
\left(\frac{r_p^2}{r_{\gamma}^2}\right)^{\gamma}\!\!
e^{\bar{\alpha}_s\chi(\gamma)Y}{\cal A}_{\mathrm{el}}(\gamma),
\label{sigmadip}
\end{eqnarray}
where $\chi_{\mathrm{LO}}(\gamma)=2\psi(1)-\psi(\gamma)-\psi(1-\gamma)$ is the BFKL kernel and
the elementary two-gluon exchange amplitude is given by ${\cal A}_{\mathrm{el}}(\gamma)=\alpha_s^2/16\gamma^2(1\!-\!\gamma)^2$. The Mellin-transform of the photon wave-function is defined by:
\begin{eqnarray}
 \int \frac{d^2r_{\gamma}}{2\pi} \int dz_{\gamma}\,\left(r_{\gamma}^2\right)^{1\!-\!\gamma}\,\Phi_{T,L}^\gamma(r_{\gamma},z_{\gamma})=\phi_{T,L}(\gamma)
  \left(Q^2\right)^{\gamma-1},\nonumber\\
 \phi_{T,L}(\gamma)=\frac{\alpha_{em}e_f^2}{\alpha_s}\frac{N_c}{4\pi}
   \frac{h_{T,L}(\gamma)}{\gamma}\;\left\{2^{-2\gamma+3}(1\!-\!\gamma)^2
   \frac{\Gamma(1\!-\!\gamma)}{\Gamma(\gamma)}\right\}\,,\nonumber
\label{eqn:phi}
\end{eqnarray}
where $h_{T,L}(\gamma)/\gamma$ is related to the Mellin transform
of the hard $\gamma^*$-gluon cross-section in the massless limit.

The final ingredient is the identification of the quantity characterizing the
average dipole size and the average number of primary dipoles. They are defined  by the following equivalence \cite{munier_peschanski}, 
\begin{eqnarray}
<\!r_p^{2\gamma}\!> \,=\int d^2 r_p \int dz_p\,\left(r_p^{2}\right)^{\gamma}\Phi^p(r_p,z_p)\equiv 
  \frac{n_{\mathrm{eff}}(\gamma)}{(Q_{\mathrm{sat}}^2)^\gamma}\ ,
\label{dip_size}
\end{eqnarray}
where $n_{\mathrm{eff}}(\gamma)$ is interpreted as the $\gamma$-dependent
average number of primary dipoles, assumed to be regular. Now, the average transverse size is not a constant value $<\!\!r_p\!\!>\simeq 2/Q_0$ but either an energy-dependent quantity, $<\!r_p\!>\,\simeq 1/Q_{\mathrm{sat}}$.

Under the latter identification, Eq. (\ref{dip_size}), and putting the additional relations together into Eq. (\ref{photoabs_xsec}) and further performing the remaining integrations, the proton structure function can be cast into the form:
\begin{eqnarray}
F_2\,(x,Q^2) & = & 32\pi^2\int\frac{d\gamma}{2i\pi}
  \left(\frac{Q^2}{Q_{\mathrm{sat}}^2}\right)^\gamma 
  e^{\bar{\alpha}_s\chi(\gamma)Y} \nonumber\\
& \times & \left[\phi_{T}(\gamma)+\phi_{L}(\gamma)\right]\,{\cal A}_{\mathrm{el}}(\gamma)\,n_{\mathrm{eff}}(\gamma)\nonumber\,.
\end{eqnarray}

The convolution integral, approximated by a steepest-descent method, using the expansion of the BFKL kernel near $\gamma=1/2$, produces the simple analytical form \cite{pheno_early,pheno_new},
\begin{eqnarray}
F_{2}(x,Q^2) & = &  {\cal N}_p\,\left(\, \frac{1}{x} \, \right)^{\omega_{\pom}}\, \left(\frac{Q^2}{Q_{\mathrm{sat}}^2}\right)^{\frac{1}{2}}\sqrt{\frac{2\,\kappa(x)}{\pi}}\nonumber \\
& \times &  \exp \left[ -\frac{\kappa\,(x)}{8}\,\ln^2 \frac{Q^2}{Q_{\mathrm{sat}}^2} \right],
\label{sfs}
\end{eqnarray}
where the overall normalization ${\cal N}$ absorbs the normalization constants and the BFKL diffusion coefficient at rapidity $Y=\ln \,(1/x)$ is written as $\kappa\,(x)=[\bar{\alpha}_s\, 7\, \zeta (3)\,Y]^{-1}$. For the saturation scale, we use the commonly considered relation $Q_{\mathrm{sat}}^2=\Lambda^2\,e^{\lambda Y}$, where we set $\lambda= 0.288$ in agreement with the saturation models \cite{GBW}. Therefore, the expression in Eq. (\ref{sfs}) with this definition for the saturation scale will be used in the phenomenological fit, where we are left with 3 free parameters: the normalization ${\cal N}_p$, the Pomeron intercept $\alpha_{\pom}=1+\omega_{\pom}$ and $\Lambda$.

\section{Results and Conclusions}

\begin{figure*}[t]
\includegraphics[scale=0.65]{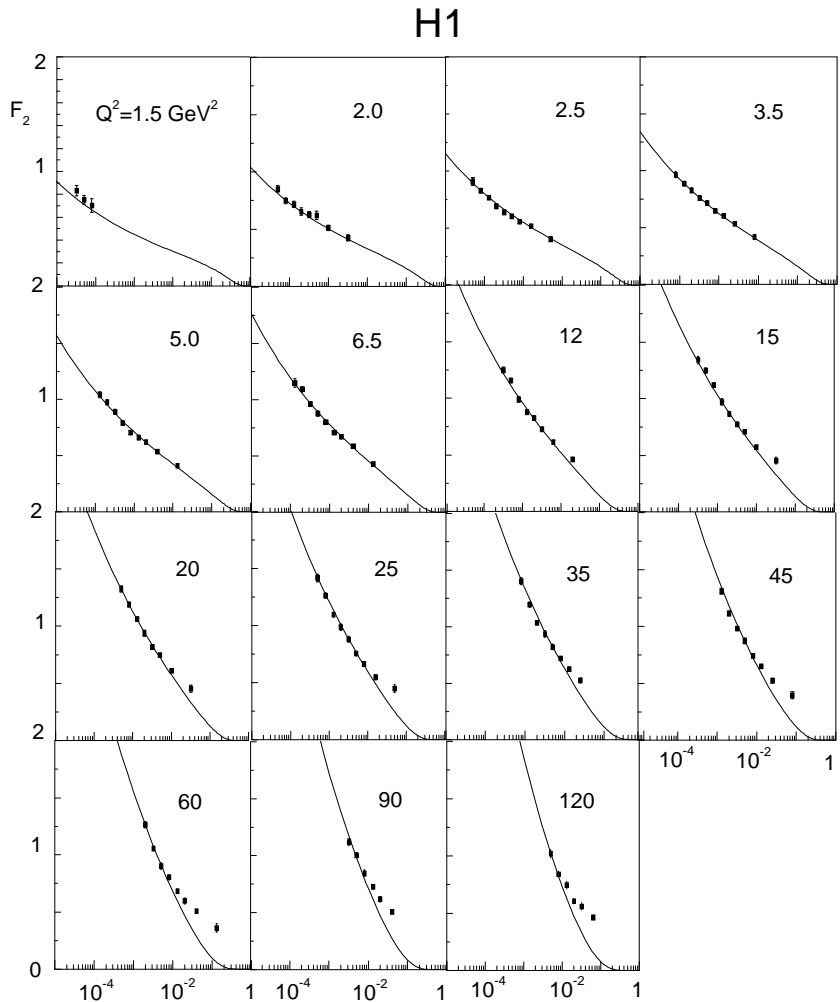}
\vspace{-3.5cm}
\caption{Proton structure function and fit result for H1 data set \cite{H1rec}. Points for larger $x>10^{-2}$ are also shown.}
\label{fig:1}
\end{figure*}

\begin{table}[t]
\caption{Parameters for H1 and ZEUS data sets \cite{H1rec,ZEUSrec,ZEUSold}.}
\label{table:1}
\begin{center}
\begin{tabular}{||c|c|c||}
\hline
\hline
 $\mathrm{PARAMETER}$  &   ZEUS data set    &  H1 data set   \\
\hline
 ${\cal N}_p$ & 0.0473  &  0.0454  \\
  $\Lambda$  & 0.119  &  0.120 \\
 $\alpha_{\pom}$ & 1.328  & 1.33 \\
\hline
\hline
$\chi^2/\mathrm{d.o.f.}$  & 1.24  &   1.28 \\
\hline
\hline
\end{tabular}
\end{center}
\end{table}

Lets present the fitting procedure using the recent DESY-HERA experimental data  on the proton structure function \cite{H1rec,ZEUSrec}, taking  Eq. (\ref{sfs})  and the small $x\leq 10^{-2}$ data. The datasets cover the ranges $0.9 \leq Q^2 \leq 90$ GeV$^2$ and $1.5\leq Q^2 \leq 120$ GeV$^2$ for ZEUS and H1, respectively. In the ZEUS case, one has added 4 bins for $0.9\!-\!2.5$ GeV$^2$ \cite{ZEUSold} since the new data set has as a lower bin $Q^2=2.7$ GeV$^2$.  The resulting parameters for H1 and ZEUS experimental data sets are presented in Table~\ref{table:1}, producing a quantitatively reasonable quality of fit in view of the high precision $F_2$ datasets. They can be compared with the recent BFKL fit in Ref. \cite{PRS} (only H1 dataset). Basically, the quality of fit is similar to the LO BFKL result in Ref. \cite{PRS}. In Fig. 1, one shows the curves using H1 parameters and the corresponding experimental measurements. We have checked the fit is not sensible to small variations in the assumed $\lambda$ value. It should be also stressed that only light quarks are considered in the approach and a treatment including charm is timely, mostly at virtualities above its production threshold. However, such a procedure would spoil the simple analytical expression considered for the fit to the proton structure function. The value obtained for $\Lambda \approx \Lambda_{\mathrm{QCD}}$ is consistent with the expectation that dipoles at initial condition $Y=0$ have average radius $\approx 1/\Lambda_{\mathrm{QCD}}$.

The procedure presented here is similar to previous analysis on Refs. \cite{pheno_early,pheno_new}, where a fixed average dipole size is considered. From those studies, one obtains a unphysical value for the effective strong coupling, $\alpha_s\simeq 0.08-0.09$.  However, with the introduction of a energy-dependent mean dipole radius the Pomeron intercept has increased ($\alpha_{\pom}=1.33$) and now the effective coupling reaches $\alpha_s\simeq 0.12$. On the other hand, next order correction are still required. Accurate analysis along these lines, considering resummed NLO BFKL kernels, has been done  \cite{PRS} with a correct running of the coupling for the typical $Q^2$ range considered in phenomenology for structure functions. Nevertheless, those NLO fits qualitatively describe the  running $\alpha_s$ effect but quantitatively the quality of fit remains sizeably higher than the LO fit (it should be stressed the number of free parameters in NLO fits is smaller). Moreover, considering the LO BFKL expression and replacing $\alpha_s$ by  running coupling $\alpha_s(Q^2)$ has also produced a low quality fit.

Finally, some comments are in order. The procedure we have used can be connected with the studies of the solution to the BFKL equation subjected to a saturation boundary condition at $Q^2\simeq Q_{\mathrm{sat}}^2$. In the saturation regime, the saturation momentum is the single intrinsic scale for hadronic processes dominated by gluons in the hadron wave function. The typical transverse size of the saturated gluons is $R_0(x)=1/Q_{\mathrm{sat}}$. It has been recently shown that the HERA data on DIS at low-$x$ are consistent with scaling in terms of the variable $\tau=Q^2R_0^2(x)$, which is known as geometric scaling phenomenon \cite{Stasto:2000er}. This scaling pattern holds outside the saturation regime and it was theoretically found \cite{Iancu:2002tr} to be extended up to $Q^2\,\lsim\,  Q_{\mathrm{sat}}^4/\Lambda_{\mathrm{QCD}}^2$. This result is obtained by realizing that the BFKL solution to the scattering amplitude (basically, the dipole cross section in Eq. (\ref{sigmadip})) is the linear limit of the Balitsky-Kovchegov (BK) evolution equation. The latter describes the high density gluons region. In fact, it was demonstrated \cite{travwaves} that geometric scaling is the exact asymptotic solution of a general class of nonlinear evolution equations \cite{KPP,BK}  and it appears as a universal property of these kind of equations. The specific scaling solutions correspond to traveling wave solutions of those equations.

Starting from the  BFKL equation  with $<\!r_p^{2}\!> \simeq 1/\Lambda^2$ and using the matching condition of its solution at saturation scale, $\sigma_{dip}(r_{\gamma }= R_0)/\sigma_0= 1$, where $\sigma_0=2\pi R_p^2$, one can study the BFKL solution near saturation momentum \cite{Iancu:2002tr}. Expanding it around the saturation scale with respect to $\ln (1/r_{\gamma}^2 \Lambda^2)$ up to the first order of the expansion one obtains $\sigma_{dip}(r_{\gamma} = 1/Q)\simeq (Q_{\mathrm{sat}}^2/Q^2)^{\gamma_{\mathrm{sat}}}$, where $\gamma_{\mathrm{sat}}$ is just a number and $x$-independent. The Pomeron intercept has been absorbed in the expansion, since it is related to the definition of the saturation scale $\ln Q_{\mathrm{sat}}^2/\Lambda^2={c\bar{\alpha}_s}\ln(1/x)$, with the coefficient $c=4\!-\!5$ determined from the saturation criterion \cite{Iancu:2002tr}. Namely, the effective Pomeron intercept, $\alpha_{\pom}=1+\omega_{\pom}$,  is related to these quantities in the form $\omega_{\pom}= c\,\bar{\alpha}_s\gamma_{\mathrm{sat}}$. In a second-order expansion near saturation scale, the BFKL solution can be written in the scaling form\cite{Iancu:2002tr},
\begin{eqnarray}
\sigma_{dip}\left(r_{\gamma}^2=\frac{1}{Q^2} \right)\simeq \left( \frac{Q_{\mathrm{sat}}^2}{Q^2}\right)^{\gamma_{\mathrm{sat}}}\exp \left[-\frac{\ln^2\left(\frac{Q^2}{Q_{\mathrm{sat}}^2}\right)}{2\,\beta\ln (1/x)}\ \right],
\label{sigdip_scal}
\end{eqnarray}
with the LO BFKL value $\beta=28 \,\zeta (3)$. The anomalous dimension at saturation limit takes the value $\gamma_{\mathrm{sat}}=0.63$, which is close to the BFKL anomalous dimension $\gamma_{\mathrm{BFKL}} \simeq 1/2$. The scaling behavior in  Eq. (\ref{sigdip_scal}) is hence transmitted to the proton structure function $F_2 \propto Q^2 \sigma_{\gamma^* p}$ with the assumption  $r_{\gamma}^2 = 1/Q^2$ (small dipole configurations in photon), namely by placing $|\Psi_{\gamma}|^2=\delta \,(r_{\gamma}^2-1/Q^2)$ in Eq. (\ref{photoabs_xsec}). On the other hand, for large dipole configurations, where dipoles on the photon have transverse size larger the saturation radius $R_0(x)$, one can suppose $|\Psi_{\gamma}|^2=\delta \,(r_{\gamma}^2-1/Q_{\mathrm{sat}}^2)$ and then the final result is a photoabsortion cross section being a constant value. This turns out the interpolation between low and intermediate $Q^2$ regions quite successful in the phenomenological applications \cite{IIM} of this approach.

Let us now return to the procedure presented here. One starts from the color dipole picture of the BFKL Pomeron and the average size of the color dipoles in the proton is given by the radius $R_0(x)$. This means the dipole density on proton somewhat  is close to the border between the dilute and the saturated limit. At the same time, we are assuming that the dipoles configurations in the photon are basically characterized by small size configurations $r_{\gamma}\ll R_0(x)$ and hence the results are valid for virtualities larger than the saturation momentum. It can be verified that the saturation criterion is automatically satisfied by construction. It remains not clear if Eq. (\ref{sfs}) can be recast in a scaling (geometric) form. However, using the identification $\omega_{\pom}= 4\ln 2\,\bar{\alpha}_s\simeq \lambda$ ($4\ln 2 \approx 2.77$), one has for the power-like rise in Eq. (\ref{sfs}), $x^{-\omega_{\pom}}\simeq x^{-\lambda}\propto Q_{\mathrm{sat}}^2$. In our case, $\gamma =1/2$ and then our expression in Eq. (\ref{sfs}) could be put in the geometric scaling form as in Eq. (\ref{sigdip_scal}).

In summary, using the  QCD dipole picture of BFKL Pomeron, one studies the role of the average dipole size in order to obtain  physical values for the effective $\alpha_s$ in  phenomenological fits to small-$x$ data. This average size is associated to the saturation scale. This is physically motivated using the assumption the color dipole on the proton have average transverse size concentrated around the saturation radius $R_0$. The quality of fit reveals this occurs up to either intermediate values of the photon virtualities.  Based on this assumption n for the mean dipole radius, the Pomeron intercept increases as $\alpha_{\pom}=1.33$) and the effective strong coupling takes a more physically acceptable value,  $\alpha_s\simeq 0.12$. However, probably NLO corrections are still required. The analysis is first performed in LO BFKL approach in the saddle-point approximation and it could be useful in phenomenology on resummed NLO BFKL kernels. 

%\vspace{-0.5cm}

\begin{acknowledgments}

%\vspace{-0.4cm}

One of us (M. Machado) is grateful for the hospitality and financial support of Service de Physique Th\'eorique, CEA/DSM/SphT, CE Saclay, where this work was accomplished. Special thanks go to Prof. Robi Peschanski for useful comments and suggestions.
\end{acknowledgments}


\begin{thebibliography}{9}
    

\bibitem{dipole} A.H. Mueller, {\it Nucl. Phys.} {\bf 415}, 373 (1994);
H. Navelet, S. Wallon, {\it Nucl. Phys.} {\bf B522}, 237 (1998).


\bibitem{BFKL}
L. N. Lipatov, {\it Sov. J. Nucl. Phys.} {\bf 23}, 338 (1976); E. A.
Kuraev, L. N. Lipatov, V. S. Fadin, {\it JETP} {\bf 45}, 1999 (1977);  I.
I. Balitskii, L. N. Lipatov, {\it Sov. J. Nucl. Phys.} {\bf 28}, 822
(1978).


\bibitem{pheno_early} H. Navelet, R. Peschanski, Ch. Royon, S.Wallon, {\it
    Phys. Lett.} {B385}, 357 (1996); 
A. Bialas, R. Peschanski, Ch. Royon, {\it Phys. Rev.} {\bf D57}, 6899 (1998); S. Munier, R.Peschanski, {\it Nucl. Phys.} {\bf B524}, 377 (1998).
    
\bibitem{pheno_new} A.I. Lengyel, M.V.T. Machado, {\it Eur. Phys. J. A.} {\bf 21}, 145 (2004). 
     
\bibitem{PRS} R. Peschanski, Ch. Royon, L. Schoeffel, arXiv:hep-ph/0411338.

\bibitem{BK}
%\bibitem{Balitsky:1995ub}
I.~Balitski\u{\i},
%``Operator expansion for high-energy scattering,''
Nucl.\ Phys.\ {\bf B463}, 99 (1996);
%%CITATION = HEP-PH 9509348;%%
Y.~V. Kovchegov,
\newblock Phys. Rev. D {\bf 60}, 034008 (1999);
%%CITATION = HEP-PH 9901281;%%
\newblock
{\bf 61}, 074018 (2000).
%%CITATION = HEP-PH 9905214;%%

\bibitem{GBW} K. Golec-Biernat and  M. W\"usthoff,  Phys. Rev. D {\bf 59}, 014017 (1999),  {\it ibid.} {\bf 60} 114023 (1999).

\bibitem{munier_peschanski} S. Munier, R. Peschanski, Nucl. Phys {\bf 524}, 377 (1998).

\bibitem{H1rec} H1 Collaboration, C. Adloff {\it et al.},  {\it Eur. Phys. J.}
{\bf C21}, 33 (2001).

\bibitem{ZEUSrec} ZEUS Collaboration, S. Chekanov {\sl et al.}, {\it Eur. Phys. J.} {\bf C21}, 443 (2001).

\bibitem{ZEUSold} ZEUS Collaboration, J. Breitweg {\sl et al.}, {\it Eur. Phys. J.} {\bf C7}, 609 (1999).

\bibitem{Stasto:2000er}
A.~M.~Sta\'sto, K.~Golec-Biernat and J.~Kwiecinski,
%``Geometric scaling for the total gamma* p cross-section in the low x
%region,''
Phys.\ Rev.\ Lett.\  {\bf 86}, 596 (2001).
%[arXiv:hep-ph/0007192].
%%CITATION = HEP-PH 0007192;%%


\bibitem{Iancu:2002tr}
E.~Iancu, K.~Itakura and L.~McLerran,
%``Geometric scaling above the saturation scale,''
Nucl.\ Phys.\ A {\bf 708}, 327 (2002); A.H. Mueller and D.N. Triantafyllopoulos,  Nucl. Phys. {\bf B640}, 331 (2002).

\bibitem{travwaves} 
S.~Munier and R.~Peschanski,
%``Geometric scaling as traveling waves,''
Phys.\ Rev.\ Lett.\  {\bf 91}, 232001 (2003).
%[arXiv:hep-ph/0309177].
%%CITATION = HEP-PH 0309177;%%


\bibitem{KPP}
R.~A. Fisher,
\newblock Ann. Eugenics {\bf 7}, 355 (1937);
A.~Kolmogorov, I.~Petrovsky, and N.~Piscounov,
\newblock Moscou Univ. Bull. Math. {\bf A1}, 1 (1937).

\bibitem{IIM} E.~Iancu, K. Itakura and S. Munier, Phys.\ Lett.\ B {\bf 590}, 199 (2004).



\end{thebibliography}
\end{document}